\documentclass[preprint,aps,showpacs]{revtex4}
\usepackage{graphicx}
\usepackage{amsmath}
\usepackage{latexsym}
\begin{document}
\preprint{Phys. Rev. E., March, 2001}
\title
{Universal scaling functions for bond percolation on \\
planar random
and square lattices with multiple \\ percolating clusters}

\author{Hsiao-Ping Hsu$^{1,*}$, Simon C. Lin$^{1,2}, $Chin-Kun Hu$^{2,3,+}$}

\affiliation
{$^1$Computing Centre, Academia Sinica, Nankang, Taipei 11529,
Taiwan \\
$^{2}$Institute of Physics, Academia Sinica, Nankang, Taipei 11529,
Taiwan \\
$^{3}$ Department of Physics, National Dong Hwa University, Hualien
97401, Taiwan}

\date{\today}
\begin{abstract}
Percolation models with multiple percolating clusters have attracted
much attention in recent years. Here
we use Monte Carlo simulations to study bond percolation on
$L_{1}\times L_{2}$ planar random lattices, duals of random lattices, and
square lattices with free and periodic boundary conditions, in vertical
and horizontal directions, respectively, and with various aspect ratio
$L_{1}/L_{2}$. We calculate the probability for the appearance of $n$
percolating clusters, $W_{n},$ the percolating probabilities, $P$, the
average fraction of lattice bonds (sites) in the percolating clusters,
$<c^{b}>_{n}$ ($<c^{s}>_{n}$)$,$ and the probability distribution function
for the fraction $c$ of lattice bonds (sites), in percolating clusters of
subgraphs with $n$ percolating clusters, $f_{n}(c^{b})$ ($f_{n}(c^{s})$).
Using a small number of nonuniversal metric factors, we find that $W_{n}$,
$P$, $<c^{b}>_{n}$ ($<c^{s}>_{n}$), and $f_{n}(c^{b})$ ($f_{n}(c^{s})$) for
random lattices, duals of random lattices, and square lattices have the same
universal finite-size scaling functions. We also find that nonuniversal
metric factors are independent of boundary conditions and aspect ratios.
\end{abstract}
\maketitle
\newpage
\section{Introduction}
Percolation is related to many interesting scientific
phenomena \cite{Stauffer94}. In recent years percolation problems
with multiple percolating clusters have attracted much attention [2-19].
Most of the simulational studies of such problems have been restricted
to percolation on lattices \cite{hw97note}. However, many
physical systems with multiple percolating clusters such as
Carbino disks used in the study of quantum Hall effects \cite{Rusin96},
or oil fields confronted with drilling problems, do not have underlined
regular lattice structures. Thus, it is of interest to know the
relationship between the quantities for percolation on regular lattices
and the quantities for percolation not on regular lattices, such
as random lattices.
In the present paper,
we use Monte Carlo simulations to study bond percolation on
$L_{1}\times L_{2}$ planar random lattices, duals of random lattices, and
square lattices with free and periodic boundary conditions in vertical
and horizontal directions, respectively, and with various aspect ratio
$L_{1}/L_{2}$. We calculate the probability for the appearance of $n$
percolating clusters, $W_{n}$, the percolating probabilities, $P$, the
average fraction of lattice bonds (sites) in percolating clusters,
$<c^{b}>_{n}$ ($<c^{s}>_{n}$), and the probability distribution function
for fraction $c$ of lattice bonds (sites), in percolating clusters of
subgraphs with $n$ percolating clusters, $f_{n}(c^{b})$ ($f_{n}(c^{s})$).
Using a small number of nonuniversal metric factors, we find that $W_{n}$,
$P$, $<c^{b}>_{n}$ ($<c^{s}>_{n}$), and $f_{n}(c^{b})$ ($f_{n}(c^{s})$) for
random lattices, duals of random lattices, and square lattices, have the same
universal finite-size scaling functions. We also find that nonuniversal
metric factors are independent of boundary conditions and aspect ratios.
Furthermore, this study is related to recent developments in the
universality and scaling of critical phenomena.

Universality and scaling are two important concepts in modern
theory of critical phenomena \cite{stanley71,Fisher71,Privman84},
and percolation models are an ideal system for studying critical phenomena
\cite {Stauffer94}. Thus, universality and scaling have been actively studied
in recent decades, especially for percolation models \cite{Okabe00}.
In 1992, Langlands
et. al. \cite{Lang92} proposed that for bond and site percolation models on
square (sq), planar triangular (pt), and honeycomb (hc) lattices, the critical
existence probability (also called crossing probability or spanning
probability) is a universal quantity, when aspect ratios of sq, pt,
and hc lattices have relative ratios, $1:\sqrt{3}:$ $\sqrt{3}/2$.
From 1995$\sim $1996, Hu, Lin and Chen (HLC) \cite{Hu95,hl96} calculated
existence probability, $E_{p}$, percolation
probability, $P$, and probability for the appearance of $n$ percolating
clusters, $W_{n},$ of bond and site percolation models,
 on sq, hc, and pt lattices with
aspect ratios, $1:\sqrt{3}:\sqrt{3}/2$, and showed that all their scaled data
fall on the same universal scaling functions, by selecting a very
small numbers of nonuniversal metric factors, and maintaining similar
nonuniversal metric
factors under two boundary conditions, free and periodic
boundary conditions.
By using renormalization group theory,
Hovi and Aharony in 1996 \cite{Hoviprl96} also pointed out that scaling
functions, for the spanning probability are universal at the fixed point for
every system with the same dimensionality, spanning rule, aspect ratio and
boundary conditions. Okabe and Kikuchi in 1996 \cite{Okabeijmp96}, extended
the work of HLC to a two-dimensional Ising model on planar regular
lattices. In 1997, Hu and Wang \cite{hw97} found that the lattice percolation
and continuum percolation of hard and soft disks, have the same
universal scaling functions for $W_n$.
Using the connection between an Ising
model and a bond-correlated percolation model \cite{hu84}, Tomita,
Okabe, and Hu in 1999 \cite{Tomita99}, calculated the probability for the
appearance of $n$ percolating clusters, $W_{n},$ the percolating
probabilities, $P$, the average fraction of lattice sites in percolating
clusters, $<c>_{n},$ and the probability distribution function for the
fraction $c$ of lattice sites in percolating clusters of subgraphs with $n$
percolating clusters, $f_{n}(c)$ , for bond-correlated percolation model
on sq, hc, and pt lattices, with aspect ratios of $1:\sqrt{3}:\sqrt{3}/2$.
Using a small number of nonuniversal metric factors, they found that $W_{n}$
, $P$, $<c>_{n},$ and $f_{n}(c)$ for sq, hc, and pt lattices have the same
universal finite-size scaling functions.

However, the studies mentioned above are mostly focused on regular lattices,
with fixed coordination numbers \cite{hw97note}.
In 1999 Hsu and Huang (HH) \cite{Hsu1999}
determined the percolation thresholds and critical exponents, and demonstrated
explicitly that the ideas of universal critical exponents and universal
scaling function with nonuniversal metric factors
can be extended to bond percolation on $L \times L$
periodic planar random lattices, duals of random lattices, and square lattices,
for both existence and percolating probabilities, and
mean cluster size. This paper will
study bond percolation on $L_1 \times L_2$ planar random lattices,
duals of random lattices, and square lattices, in
more detail, and consider the case that the lattices have
free and periodic boundary conditions, in vertical and horizontal
directions, respectively, as in \cite{hl96}.
Percolating probability is defined by the
ratio of the number of bonds in the percolating clusters to the total number
of bonds in \cite{Hsu1999}. Here, we consider two different definitions
of the percolating probability, in terms of bonds and in terms of sites;
the latter was also used in \cite{hl96} and \cite{Tomita99}. We calculate
the probability $W_{n}$ for the appearance of $n$ percolating
clusters, the percolating probability $P$, the average fraction of lattice
bonds (sites) in percolating clusters, $<c^{b}>_{n}$ $(<c^{s}>_{n})$,
and the probability distribution function for the fraction $c$ of lattice
bonds (sites) in percolating clusters of subgraphs with $n$ percolating
clusters, $f_{n}(c^{b})$ $(f_{n}(c^{s}))$, for
various values of aspect ratios $L_1 /L_2$, and finally check
the universal finite-size scaling behaviors for these quantities.
In \cite{Hsu1999}, HH used two nonuniversal metric factors,
$D_{2}$ and $D_{3}$, to fix universal finite-size scaling functions,
for percolating probability in
terms of bonds. In the present paper, we calculated
two nonuniversal metric factors of
percolating probability, in terms of sites and obtained previous known values
of nonuniversal metric factors determined by HH,
to check whether we could have universal scaling
functions for $W_{n}$, $P$, $<c^{b}>_{n}$ $(<c^{s}>_{n})$, and $f_{n}(c^{b})$
$(f_{n}(c^{s}))$, of bond percolation on random lattices, duals of random
lattices, and square lattices.

Dirichlet and Voronoi \cite{Itzykson92} first used the concept of random
lattices in condensed matter theory and Christ, Friedberg, and Lee (CFL)
\cite{Christ82a} used another type of random lattices to formulate quantum
field theory. Here, we adopt the CFL algorithm and give a brief review of the
construction of planar random lattices and their duals.
First, we randomly generate $N$ sites in the $L_1 \times L_2$ rectangular
domain with periodic boundary conditions. Next, we choose three nearby
sites arbitrarily, and draw a circle to go through the three sites. If there
are no lattice sites inside the circle, the three sites are connected by
links to form a triangle. A planar random lattice is constructed by
repeating the process until all sites are connected by links. The whole
rectangular domain is divided into $2N$ non-overlapping triangles,
whose vertices are sites of the random lattice, and circle centers
with triangles are the sites of dual lattice.
Thus, there is a one to one correspondence between
triangles and dual lattice sites. Because a link
of the random lattice is shared by two
triangles, the two corresponding dual lattice sites are connected by one dual
link. There is a one to one correspondence between links and dual links.
The whole rectangular domain is partitioned into $N$ non-overlapping planar
convex polyhedra, which are formed by dual links and the vertices of $N$
polyhedra are sites for dual lattice. There is also a one to one
correspondence between the lattice sites and polyhedra on dual lattice.
Examples of a planar random lattice with its dual, under periodic boundary
conditions, in both vertical and horizontal directions, are shown in Fig. 1(a).

This paper is organized as follows: In Sec. II, we present the simulational
results for $W_{n}$, $P$, $<c^{b}>_{n}$ ($<c^{s}>_{n}$),
and $f_{n}(c^{b})$ ($f_{n}(c^{s})$) for bond percolation, on
$L_1 \times L_2$
random lattices, duals of random lattices, and square lattices,
under free and periodic boundary conditions in vertical and horizontal
directions with $L_1/L_2=4$.
The boundary bonds which cross the rectangular domain in
vertical direction on random lattices, due to periodic boundary
conditions, are eliminated because
of free boundary conditions in vertical direction
considered in this paper.
We adopt the method of HH \cite{Hsu1999}, to find percolating clusters.
Only the first kind of percolating cluster paths, without boundary
bonds in the vertical direction (the clusters extend from top to bottom),
should be identified, and an example of this is shown in Fig. 1(b).
In Sec. III, we use finite-size
scaling theory to check the scaling behaviors of various quantities,
and to show that such quantities have universal finite-size
scaling functions for regular lattices and random lattices.
A summary is provided in Sec. IV.

\section{$W_{\lowercase{n}}(L_{1},L_{2},\lowercase {p})$,
\lowercase{$f_{n}(c)$, and $<c>_n$}}

We see the bond percolation on a lattice $G$, with linear dimensions $L_{1}$
and $L_{2}$ in horizontal and vertical directions, respectively; the
probability for the appearance of $n$ top-to-bottom percolating clusters,
$W_{n}(L_{1},L_{2},p),$ is defined by \cite{hl96}
\begin{equation}
W_{n}(L_{1},L_{2},p)=\sum\limits_{G_{n}^{\prime }\subseteq
G}p^{b(G_{n}^{\prime })}(1-p)^{E-b(G_{n}^{\prime })}.
\end{equation}
Here, the percolating cluster is defined as a cluster extending from top to
bottom in $G$, $G_{n}^{\prime }$ denotes a percolating subgraph with $n$
percolating clusters, $b(G_{n}^{\prime })$ is the number of occupied bonds
in $G_{n}^{\prime }$, and $E$ is the total number of links in $G$. The
existence probability $E_{p}$ can be expressed obviously as
\begin{equation}
E_{p}=\sum\limits_{n=1}^{\infty }W_{n},
\end{equation}
with $W_{0}=1-E_{p}$.

To obtain more detailed information about the contents of the percolating
cluster, following Tomita et. al. \cite{Tomita99}, we decompose $W_{n}$ as
\begin{equation}
W_{n}=\int\nolimits_{0}^{1}f_{n}(c)dc,
\end{equation}
where $n=1,\ldots \infty $, $c$ denotes the fraction of lattice bonds
(sites) in percolating clusters, and $f_{n}(c)$ is the probability
distribution function of $c$ in subgraphs with $n$ percolating clusters. The
probability distribution function of $c$ in all subgraphs is the overall
summations of $f_{n}(c),$ i.e.,
\begin{equation}
f(c)=\sum\limits_{n=1}^{\infty }f_{n}(c).
\end{equation}
In terms of $f_{n}(c)$, the average fraction of lattice bonds (sites) in
subgraphs with $n$ percolating clusters can be expressed as
\begin{equation}
<c>_{n}=\int\nolimits_{0}^{1}cf_{n}(c)dc,
\end{equation}
where $n=1,\ldots ,\infty ,$ and percolating probability $P$ can be
written as
\begin{equation}
<c>=\sum\limits_{n=1}^{\infty }<c>_{n}=\int_0^{\infty}cf(c)dc=P.
\end{equation}

To generate subgraphs, we use the random bond occupation process with equal
occupation probability for each link. The simulations are
performed on $128\times 32$, $256\times 64$, and $512\times 128$ planar
random (pran) lattices, and their duals (dpran), with free and periodic
boundary conditions in the vertical and horizontal directions, respectively.
To compare the results with regular lattices, we also perform simulations
on square (sq) lattices of the same sizes.
On each lattice, we take
$60$ occupation probabilities around the critical percolation threshold for
every $0.002$ increment, and use the random bond occupation process
to generate $10^{5}\sim 10^{6}$ configurations
 for each occupied probability, $p$. We
calculate $W_{n}(L_{1},L_{2},p)$ , $<c^{b}>_{n}$, and $<c^{s}>_{n}$, where
$c^{b}$ denotes the fraction of bonds in percolating clusters, and $c^{s}$
denotes the fraction of sites in percolating clusters; and the results are
shown in Figs. 2 and 3. The calculated results of the percolating
probabilities in terms of bonds, $P^{b},$ and in terms of sites, $P^{s}$, are
also shown in Fig. 3. We calculate $f_{n}(c^{b})$ and $f_{n}(c^{s})$ at
$p=p_{c}$ and take $p_{c}=0.3333$ for planar random lattices, and
$p_{c}=0.6667$
for dual lattice \cite{Hsu1999}. The results are shown in Fig. 4.
The differences between bond and site contents in percolating clusters are
shown in Figs. 3 and 4; here, for the clarity of presentation only, the
results for $512\times 128$ lattices are plotted in the figures.

\section{Universal Finite-size Scaling Functions}
The finite-size scaling theory was first formulated by Fisher in 1971 \cite
{Fisher71}. According to the theory, for a physical quantity $X,$
which scales as $X(t)\sim t^{\rho }$ in a thermodynamic system near
critical point $t=0$, then the same quantity in a finite system with
linear dimension $L$, $X_L(t)$, should obey the general law,
\begin{equation}
X_{L}(t)\sim L^{-\rho /\nu }F(tL^{1/\nu }).
\end{equation}
Here, $F(x)$ with $x=tL^{1/\nu }$ labelled as scaling function, with $\nu $
as correlation length exponent. In 1984, Privman and Fisher \cite{Privman84}
considered universal finite-size
scaling functions and nonuniversal metric factors,
and proposed that the singular part of the free
energy of a critical system can be written as
\begin{equation}
f_{L}^{s}(t)\sim L^{-d}Y(DtL^{1/\nu }),
\end{equation}
where $d$ is the spatial dimensionality of lattice, $Y$ is a universal
scaling function, and $D$ is a nonuniversal metric factor.

At the critical point, $p=p_{c}$, there also exists a finite-size scaling
form for the distribution function of $X_{L}(t)$ \cite{Tomita99}:
\begin{equation}
Q(X_L(t=0))\sim L^{\rho /\nu }Y(X_L(t=0)\cdot L^{\rho /\nu }).
\end{equation}
In \cite{Hu95,Hsu1999}, three
nonuniversal metric factors $D_{1}$, $D_{2}$, and $D_{3}$ were used for
regular lattices and
random lattices, to describe the universal scaling functions of
existence probability $E_{p}$, and the percolating probability $P$, i.e.
\begin{equation}
E_{p}(p,L)=F(x),
\end{equation}
with $x=D_{1}(p-p_{c)}L^{1/\nu }$ and
\begin{equation}
D_{3}P(p,L)=L^{-\beta /\nu }S_{p}(z),
\end{equation}
with $z=D_{2}(p-p_{c})L^{1/\nu }.$

Following Hu et. al. \cite{hl96}, we use the evaluated percolation
thresholds $p_{c}$ \cite{Hsu1999}, and the exact values of critical exponent
$\nu =4/3$, to plot $W_{n}$ as a function of $y=(p-p_{c})L^{1/\nu }$, for
planar random lattices and their duals in
Figs. 5(a) and 5(b), respectively.
We can see from these results, that the scaled data for $W_{n}$ can be
described by a single scaling function $F_{n}(r,y)$ with $r=L_{1}/L_{2}$, and
$F_{n}(r,y)$ for $n\geq 2$ as a symmetric function of $y$.
In Fig. 6 we plot
$W_{n}(L_{1},L_{2},p)$ as a function of $x$ for bond percolation on
$512\times 128$ random lattice, dual of random lattice, and square lattice,
where $x=D_{1}(p-p_{c})L^{1/\nu }$, with $D_{1}$ taken from Table I and
$L=(L_{1}\times L_{2})^{1/2}$. Fig. 6 shows that the calculated
results for each $n$ can be well described by a single universal scaling
function, $U_{n}(x)$.

In Hu and Lin's paper \cite{hl96}, the scaling functions $F_{n}(r,y)$
were calculated for bond percolation on
square lattice for various values of aspect ratios, $r$.
We will examine whether the same nonuniversal metric factors, $D_{1}$, can be
extended to different aspect ratios. We calculate $W_{n}$ for $L_{1}\times
L_{2}$ random lattices, dual of random lattices, and square lattices with
$r=L_{1}/L_{2}=1,$ $2,$ $\ldots ,$ $6$ and determine the universal scaling
functions, $U_{n}(r,x),$ where $x=D_{1}(p-p_{c})L^{1/\nu }$ and where $D_{1}$
is taken from Table I. The results for $n=1$ and $2$ are shown in
Figs. 7(a) and 7(b), respectively. We can see that the scaled data for
each $r$ can be described by a single universal scaling function very well.
The results of $U_{n}(r,x)$ as a function of $r$ for
$n=0,1,\ldots ,$ $4$ at
the critical point $p=p_{c}$, are presented in Fig. 8(a) which show
that the three lattices provide similar results.
We also calculate
the average number of percolating clusters $C(r,x)$, defined by
\begin{equation}
C(r,x)=\sum\limits_{n=1}^{\infty }U_{n}(r,x)n.
\end{equation}
$C(r,0)$ for random lattice, dual of random lattice, and square lattice as a
function of $r$ are shown in Fig. 8(b). Fig. 8(b) shows that $C(r,0)$
increases linearly with an increasingly large, $r$ and that different lattices
have the same slope of approximately 0.43.

Tomita et. al. \cite{Tomita99} had obtained universal finite-size scaling
functions of $<c>_{n}$, $<c>,$ $f_{n}(c),$ and $f(c)$ for a bond-correlated
percolation model, corresponding to an Ising model on planar regular
lattices. It is of interest to extend such study to bond random percolation
on random lattices. From Eqs.(5), (6), and (11), the universal scaling
function of $<c>$ and $<c>_{n}$ can be expressed as
\begin{equation}
D_{3}<c(p,L)>=L^{-\beta /\nu }G(z),
\end{equation}
and
\begin{equation}
D_{3}<c(p,L)>_{n}=L^{-\beta /\nu }G_{n}(z),
\end{equation}
with $z=D_{2}(p-p_{c})L^{1/\nu }.$ At $p=p_{c},$ the universal scaling
function of $f(c)$ and $f_{n}(c)$ are expressed as
\begin{equation}
D_{3}^{-1}f(c)=L^{\beta /\nu }H(z^{\prime }),
\end{equation}
and
\begin{equation}
D_{3}^{-1}f_{n}(c)=L^{\beta /\nu }H_{n}(z^{\prime }),
\end{equation}
with $z^{\prime }=D_{3}cL^{\beta /\nu }.$ To check the finite-size scaling
and universality of these quantities, we use simulation results for
$256\times 64$ and $512\times 128$ square lattices, planar random lattices
and their duals. In \cite{Hsu1999}, the percolating probability $P$ is
defined in terms of the bond number in percolating clusters, and the
nonuniversal metric factors $D_{2}=D_{2}^{b}$ and $D_{3}=D_{3}^{b}$ are
used. To evaluate factors $D_{2}^{s}$ and $D_{3}^{s}$, we adopt the same
procedure as in \cite{Hsu1999}, plotting $P^{b}/L^{-\beta /\nu }$ and
$P^{s}/L^{-\beta /\nu }$ as functions of $y=(p-p_{c})L^{1/\nu }$, as
shown in Fig. 9.
All of the nonuniversal metric factors for different
types of lattices used in this paper are listed in Table I.

We plot $D_{3}P/L^{-\beta /\nu }$ and $D_{3}<c>_{n}/L^{-\beta /\nu }$ as a
function of $z=D_{2}(p-p_{c})L^{1/\nu }$, in Figs. 10(a) and 10(b) for bond
content and site content, respectively, with $D_{2}$ and $D_{3}$ taken from
Table I. At $p=p_{c},$ the scaled data $D_{3}^{-1}f(c)/L^{\beta /\nu
}$, and $D_{3}^{-1}f_{n}(c)/L^{\beta /\nu }$ as functions of $z^{\prime
}=D_{3}cL^{\beta /\nu }$, are presented in Figs. 11(a) and 11(b),
respectively, for the bond content and site content. Figs. 10 and 11 show that
the bond percolation processes on square lattices, random lattices and their
duals have universal finite-size scaling functions.

\section{Summary and Discussion}
Having used nonuniversal metric factors from Table I in this
paper, we have found that the universal finite-size scaling functions
for $W_n$ (Figs. 6 and 7), $<c^b>_n$ and $P^b$ (Fig. 10(a)),
$<c^s>_n$ and $P^s$ (Fig. 10(b)), $f_{n}(c^b)$ (Fig. 11(a)),
and $f_{n}(c^s)$ (Fig. 11(b)). Fig. 7 includes results for different
aspect ratios $r$, i.e. $6 \geq r \geq 1$. The values of
nonuniversal metric factors, $D_1$, $D_2^b$, and $D_3^b$ of Table I
are consistent with the corresponding values of
Ref. \cite{Hsu1999}, where the boundary conditions are different from
the boundary conditions of the present paper. These results suggest
that in random lattices, the nonuniversal metric factors are also
independent of the boundary conditions and aspect ratios, as in the
case of regular lattices \cite{Hu95}. Please also note that
$D_2^s$ and $D_3^s$ in Table I are consistent within numerical errors.

Many interesting problems are related to the properties of multiple
percolating clusters. It is of interest to extend the study of the
present paper to higher spatial dimensions. In particular, the further study
of multiple percolating clusters in three dimensions could be
related to an oil drilling problem.

\begin{acknowledgments}
This work was supported in part by the National Science Council of the
Republic of China (Taiwan) under Contract No. NSC 89-2112-M-001-084.
\end{acknowledgments}

\newpage
Table~ I: The values of metric factors $D_{1}$, $D_{2}^{b}$, $D_{3}^{b}$,
$D_{2}^{s}$ and $D_{3}^{s}$, for square lattices, random lattices and their
duals, with free and periodic boundary conditions in vertical and
horizontal directions, respectively.\vskip0.2 true cm
\begin{tabular}{cccccccccc}
\hline\hline
Lattices    &&& Square &&& Planar random &&& Dual of planar random \\ \hline
$D_{1}$     &&& $1$    &&& $1.166\pm 0.020$ &&& $1.177\pm 0.016$ \\
$D_{2}^{b}$ &&& $1$    &&& $1.164\pm 0.014$ &&& $1.176\pm 0.015$ \\
$D_{3}^{b}$ &&& $1$    &&& $1.512\pm 0.008$ &&& $0.778\pm 0.002$ \\
$D_{2}^{s}$ &&& $1$    &&& $1.186\pm 0.012$ &&& $1.180\pm 0.014$ \\
$D_{3}^{s}$ &&& $1$    &&& $1.062\pm 0.001$ &&& $1.005\pm 0.002$ \\ \hline
\end{tabular}
\newpage

\noindent
{\bf FIGURES}
\vskip 1.0 cm

\noindent
FIG.~1. Examples of (a) $L_{1}\times L_{2}=8\times 4$ planar random lattice
(solid lines), with its dual (dashed lines) on $L_{1}\times L_{2}=8\times 4$
rectangular area, with periodic boundary conditions, and (b) a first kind of
percolating cluster path, without boundary bonds (bold solid lines) on
random lattice.
\vskip 0.5 cm
\noindent
FIG.~2. $W_{n}(L_{1},L_{2},p)$ on square lattices, planar random lattices
and their duals of size $128\times 32$, $256\times 64$, and $512\times 128$,
which are represented by dotted, dashed, and solid lines, respectively.
\vskip 0.5 cm

\noindent
FIG.~3. (a) $P^{b}$ and $<c^{b}>_{n},$ and (b) $P^{s}$ and $<c^{s}>_{n}$
on square lattice, planar random lattice and its dual of same size
$512\times 128$.
\vskip 0.5 cm

\noindent
FIG.~4. At $p=p_{c}$, (a) $f(c^{b})$ and $f_{n}(c^{b}),$ and (b) $f(c^{s})$
and $f_{n}(c^{s})$ on square lattice, planar random lattice and its dual
of same size $512\times 128$.
\vskip 0.5 cm

\noindent
FIG.~5. The scaled results of $F_{n}(r,y)=$ $W_{n}(L_{1,}L_{2},p)$ as a
function of $y=(p-p_{c})L^{1/\nu }$ for (a) planar random lattices and (b)
their duals of size $128\times 32$, $256\times 64$, and $512\times 128.$
The monotonic decreasing function is for $F_{0}(r,y).$ The S shaped curve is
for $F_{1}(r,y)$. \ The bell shaped curves from top to bottom are for
$F_{n}(r,y)$, with $n$ as $2$, $3$, and $4$, respectively.
\vskip 0.5 cm

\noindent
FIG.~6. The scaled results of $U_{n}(x)=$ $W_{n}(L_{1,}L_{2},p)$, as a
function of $x=D_{1}(p-p_{c})L^{1/\nu }$ for square lattice (solid curves),
planar random lattice (dashed curves), and the dual of planar random lattice
(dotted curves) of same size $512\times 128.$
\vskip 0.5 cm

\noindent
FIG.~7. $U_{n}(r,x)$ for square lattice (solid curves), planar random
lattice (dashed curves), and the dual of planar random lattice (dotted curves).
(a) $n=1,$ the intersection of the curves on $x=0$ axis, from up to down are
for $r=L_{1}/L_{2}=1,2,3,\ldots ,6.$ (b) The width of curves, from small
to large are for $r=L_{1}/L_{2}=1,2,3,\ldots ,6.$
\vskip 0.5 cm

\noindent
FIG.~8. (a) $U_{n}(r,0)$ as a function of $r=L_{1}/L_{2}$, for a number of
percolating clusters (npc) run from $0$ to $4$, and (b) $C(r,0)$ as a
function of $r=L_{1}/L_{2}$, with slope of the fitting line $0.43$.
Square lattice $(\Box),$ planar random lattice $(\triangle)$ and
the dual of planar random lattice $(\times)$, all have horizontal periodic
boundary conditions.
\vskip 0.5 cm

\noindent
FIG.~9. The scaled results of $P(p,L)/L^{-\beta /\nu }$, in terms of bonds
$(P=P^{b})$ and sites $(P=P^{s})$, for different types of lattice size
$256\times 64$ and $512\times 128$, as a function of $y=(p-p_{c})L^{1/\nu}$.
\vskip 0.5 cm

\noindent
FIG.~10. The scaled results of
$G_{n}(z,L)=$ $D_{3}<c(p,L)>_{n}/L^{-\beta /\nu}$,
as a function of $z=D_{2}(p-p_{c})L^{1/\nu }$, for square lattices,
planar random lattices and their duals of same size $256\times 64$,
and $512\times 128$. (a) $c=c^{b}$, $D_2=D_2^b$, $D_3=D_3^b$, and (b)
$c=c^{s}$, $D_2=D_2^s$, $D_3=D_3^s$. \vskip 0.5 cm

\noindent
FIG.~11. The scaled results of
$H_{n}(z^{\prime},L)=D_{3}^{-1}f_{n}(c)/L^{\beta /\nu }$,
as a function of $z^{\prime }=D_{3}(p-p_{c})L^{\beta /\nu }$ for square
lattices, planar random lattices, their duals of same size $256\times
64$, and $512\times 128$, with fitting results (solid curves);
(a) $c=c^{b}$, $D_{3}=D_{3}^{b}$, and (b) $c=c^{s}$, $D_{3}=D_{3}^{s}$.

\begin{thebibliography}{99}
\bibitem{hw97note} One exception is the study of continuum percolation
of soft disks and hard disks, see \cite{hw97}.
\end{thebibliography}
\end{document}